\documentstyle[twoside,fleqn,epsfig,axodraw,npb]{article}

\newcommand{\la}[1]{\label{#1}}
\newcommand{\be}{\begin{equation}}
\newcommand{\ee}{\end{equation}}
\newcommand{\ba}{\begin{eqnarray}}
\newcommand{\ea}{\end{eqnarray}}
\newcommand{\fig}{Fig.~$\!$}
\newcommand{\figs}{Figs.~$\!$}
\newcommand{\eq}{Eq.~}
\newcommand{\nr}[1]{(\ref{#1})}
\newcommand{\nn}{\nonumber \\}
\newcommand{\fr}[2]{{\frac{#1}{#2}\,}}
\renewcommand{\(}{\left(}
\renewcommand{\)}{\right)}
\newcommand{\lb}{\left\{}
\newcommand{\rb}{\right\}}

\newcommand{\e}{\epsilon}
\renewcommand{\l}{\ell}
\newcommand{\6}{\partial}

\def\Asc(#1,#2)(#3,#4,#5){\CArc(#1,#2)(#3,#4,#5)}
\def\Lsc(#1,#2)(#3,#4){\Line(#1,#2)(#3,#4)}
\def\Ahh(#1,#2)(#3,#4,#5){\DashCArc(#1,#2)(#3,#4,#5){1}}
\def\Lhh(#1,#2)(#3,#4){\DashLine(#1,#2)(#3,#4){1}}
\def\Aqu(#1,#2)(#3,#4,#5){\ArrowArc(#1,#2)(#3,#4,#5)}
\def\Aaqu(#1,#2)(#3,#4,#5){\ArrowArcn(#1,#2)(#3,#5,#4)}
\def\scfc{0.7}  
\def\phgt{21}   
\def\pwc{21}    
\def\pwcb{31.5} 
\newcommand{\PIC}[4]{\;\parbox[c]{#2 pt}{\begin{picture}(#2,#3)(0,0)
\SetWidth{1.0}\SetScale{#4} #1 \end{picture}}\;}
\newcommand{\pic}[1]{\PIC{#1}{\pwc}{\phgt}{\scfc}}
\newcommand{\picb}[1]{\PIC{#1}{\pwcb}{\phgt}{\scfc}}
\def\TopoVR(#1){\pic{#1(15,15)(15,-90,270)}}
\def\ToptVS(#1,#2,#3){\pic{#1(15,15)(15,0,180) #2(15,15)(15,180,360)%
 #3(30,15)(0,15)}}
\def\ToprVM(#1,#2,#3,#4,#5,#6){\pic{#3(15,15)(15,-30,90) #1(15,15)(15,90,210)%
 #2(15,15)(15,210,330) #5(2,7.5)(15,15) #6(15,15)(15,30) #4(28,7.5)(15,15)}}
\def\ToprVV(#1,#2,#3,#4,#5){\!\!\picb{#2(26.25,15)(15,256,76)%
 #3(30,30)(15,30) #1(18.75,15)(15,104,284) #4(15,30)(22.5,0)%
 #5(30,30)(22.5,0)}\!\!}
\def\ToprVB(#1,#2,#3,#4){\picb{#1(30,15)(15,-120,120) #2(30,15)(15,120,240)%
 #3(15,15)(15,60,300) #4(15,15)(15,-60,60)}}
\def\TopfVX(#1,#2,#3,#4,#5,#6,#7,#8,#9){\picb{#1(15,15)(15,90,270)%
 #2(30,15)(15,-90,90) #4(30,30)(15,30) #3(15,0)(30,0) #6(15,0)(15,15)%
 #5(15,15)(30,30) #8(15,30)(20,25) #8(25,20)(30,15) #7(30,15)(30,0)%
 #9(15,15)(30,15)}}
\def\TopfVH(#1,#2,#3,#4,#5,#6,#7,#8,#9){\picb{#1(15,15)(15,90,270)%
 #2(30,15)(15,-90,90) #4(30,30)(15,30) #3(15,0)(30,0) #6(15,0)(15,15)%
 #5(15,15)(15,30) #8(30,30)(30,15) #7(30,15)(30,0) #9(15,15)(30,15)}}
\def\TopfVW(#1,#2,#3,#4,#5,#6,#7,#8){\pic{#1(15,15)(15,90,180)%
 #3(15,15)(15,180,270) #2(15,15)(15,270,360) #4(15,15)(15,0,90)%
 #5(15,15)(15,30) #7(15,15)(15,0) #6(0,15)(15,15) #8(30,15)(15,15)}}
\def\TopfVWdot(#1,#2,#3,#4,#5,#6,#7,#8){\pic{#1(15,15)(15,90,180)%
 #3(15,15)(15,180,270) #2(15,15)(15,270,360) #4(15,15)(15,0,90)%
 #5(15,15)(15,30) #7(15,15)(15,0) #6(0,15)(15,15) #8(30,15)(15,15)%
 \Vertex(22,15){2}}}
\def\TopfVWlap(#1,#2,#3,#4,#5,#6,#7,#8){\pic{#1(15,15)(15,90,180)%
 #3(15,15)(15,180,270) #2(15,15)(15,270,360) #4(15,15)(15,0,90)%
 #5(15,15)(15,30) #7(15,15)(15,0) #6(0,15)(15,15) #8(30,15)(15,15)%
 \Text(2,19)[br]{$\scriptstyle 1$}\Text(20,2)[tl]{$\scriptstyle 2$}}}
\def\TopfVV(#1,#2,#3,#4,#5,#6,#7,#8){\!\!\picb{#2(26.25,15)(15,256,346)%
 #3(26.25,15)(15,-14,76) #4(30,30)(15,30) #1(18.75,15)(15,104,284)%
 #7(22.5,0)(15,30) #6(30,30)(26.25,15) #8(26.25,15)(22.5,0)%
 #5(25.25,15)(39.8,11.4)}\!\!}
\def\TopfVB(#1,#2,#3,#4,#5,#6,#7){\picb{#2(30,15)(15,-120,120)%
 #6(30,15)(15,120,180) #5(30,15)(15,180,240) #1(15,15)(15,60,300)%
 #4(15,15)(15,-60,0) #3(15,15)(15,0,60) #7(30,15)(15,15)}}
\def\TopfVBdot(#1,#2,#3,#4,#5,#6,#7){\picb{#2(30,15)(15,-120,120)%
 #6(30,15)(15,120,180) #5(30,15)(15,180,240) #1(15,15)(15,60,300)%
 #4(15,15)(15,-60,0) #3(15,15)(15,0,60) #7(30,15)(15,15)%
 \Vertex(28,22){2}}}
\def\TopfVBlap(#1,#2,#3,#4,#5,#6,#7){\picb{#2(30,15)(15,-120,120)%
 #6(30,15)(15,120,180) #5(30,15)(15,180,240) #1(15,15)(15,60,300)%
 #4(15,15)(15,-60,0) #3(15,15)(15,0,60) #7(30,15)(15,15)%
 \Text(9,15)[r]{$\scriptstyle 1$}\Text(22.5,15)[l]{$\scriptstyle 2$}}}
\def\TopfVN(#1,#2,#3,#4,#5,#6,#7){\picb{#1(15,15)(15,90,270)%
 #2(30,15)(15,-90,90) #4(30,30)(15,30) #3(15,0)(30,0)%
 #5(15,0)(15,30) #6(30,30)(30,0) #7(15,30)(30,0)}} 
\def\TopfVU(#1,#2,#3,#4,#5,#6,#7){\pic{#3(15,15)(15,0,90)%
 #2(15,15)(15,90,180) #4(15,15)(15,180,270) #1(15,15)(15,270,360)%
 #6(0,15)(15,30) #7(15,0)(0,15) #5(30,15)(15,0)}}
\def\TopfVT(#1,#2,#3,#4,#5,#6){\pic{#1(15,15)(15,90,210)%
 #2(15,15)(15,210,330) #3(15,15)(15,-30,90) #4(2,7.5)(15,30)%
 #6(28,7.5)(2,7.5) #5(15,30)(28,7.5)}}
\def\TopLV(#1,#2,#3,#4,#5,#6){\!\!\picb{#2(26.25,15)(15.5,256,76)%
 #3(30,30)(15,30) #1(18.75,15)(15.5,104,284) #4(15,30)(22.5,0)%
 #5(30,30)(22.5,0) #6(15,17.8)(19.3,292.8,39.1)}\!\!}
\def\TopfVBB(#1,#2,#3,#4,#5){\picb{#1(30,15)(15,-120,120)%
 #2(30,15)(15,120,240) #3(15,15)(15,60,300) #4(15,15)(15,-60,60)%
 #5(22.5,3)(22.5,27)}}
\def\TopfVBBdot(#1,#2,#3,#4,#5){\picb{#1(30,15)(15,-120,120)%
 #2(30,15)(15,120,240) #3(15,15)(15,60,300) #4(15,15)(15,-60,60)%
 #5(22.5,3)(22.5,27) \Vertex(22.5,10){2} \Vertex(22.5,20){2}}}
\def\TopfVBBlap(#1,#2,#3,#4,#5){\picb{#1(30,15)(15,-120,120)%
 #2(30,15)(15,120,240) #3(15,15)(15,60,300) #4(15,15)(15,-60,60)%
 #5(22.5,3)(22.5,27)%
 \Text(2,10.5)[l]{$\scriptstyle 1$}\Text(29.5,10.5)[r]{$\scriptstyle 2$}}}

\def\one{\TopoVR(\Asc)}
\def\two{\ToptVS(\Asc,\Asc,\Lsc)}
\def\threeM{\ToprVM(\Asc,\Asc,\Asc,\Lsc,\Lsc,\Lsc)}
\def\threeV{\ToprVV(\Asc,\Asc,\Lsc,\Lsc,\Lsc)}
\def\threeB{\ToprVB(\Asc,\Asc,\Asc,\Asc)} 
\def\topoEX{\TopfVX(\Asc,\Asc,\Lsc,\Lsc,\Lsc,\Lsc,\Lsc,\Lsc,\Lsc)} 
\def\topoI{\TopfVH(\Asc,\Asc,\Lsc,\Lsc,\Lsc,\Lsc,\Lsc,\Lsc,\Lsc)} 
\def\topoII{\TopfVW(\Asc,\Asc,\Asc,\Asc,\Lsc,\Lsc,\Lsc,\Lsc)} 
\def\topoIII{\TopfVV(\Asc,\Asc,\Asc,\Lsc,\Lsc,\Lsc,\Lsc,\Lsc)}
\def\topoIV{\TopfVB(\Asc,\Asc,\Asc,\Asc,\Asc,\Asc,\Lsc)}
\def\topoV{\TopfVN(\Asc,\Asc,\Lsc,\Lsc,\Lsc,\Lsc,\Lsc)} 
\def\topoVI{\TopfVU(\Asc,\Asc,\Asc,\Asc,\Lsc,\Lsc,\Lsc)}
\def\topoVIII{\TopfVT(\Asc,\Asc,\Asc,\Lsc,\Lsc,\Lsc)} 
\def\topoIX{\TopLV(\Asc,\Asc,\Lsc,\Lsc,\Lsc,\Asc)} 
\def\topoXII{\TopfVBB(\Asc,\Asc,\Asc,\Asc,\Lsc)} 

\def\topoXI{\one\threeV}

\def\topoXIV{\(\!\!\one\!\!\)^2\!\!\two}
\def\topoXV{\(\!\!\one\!\!\)^4}
\def\threeQED{\ToprVB(\Asc,\Asc,\Ahh,\Ahh)}

\def\caseXIV{\TopfVBB(\Asc,\Ahh,\Asc,\Ahh,\Lhh)}
\def\caseXIII{\TopfVBB(\Asc,\Asc,\Asc,\Asc,\Lhh)}
\def\caseX{\TopfVT(\Ahh,\Ahh,\Ahh,\Lsc,\Lsc,\Lsc)}
\def\caseIX{\TopfVT(\Ahh,\Asc,\Ahh,\Lhh,\Lhh,\Lsc)}
\def\caseVIII{\TopfVT(\Ahh,\Asc,\Asc,\Lhh,\Lsc,\Lsc)}
\def\caseVII{\TopfVT(\Asc,\Asc,\Asc,\Lsc,\Lsc,\Lsc)}
\def\caseIV{\TopfVB(\Asc,\Asc,\Asc,\Ahh,\Ahh,\Asc,\Lsc)}
\def\caseIII{\TopfVB(\Ahh,\Asc,\Asc,\Ahh,\Asc,\Ahh,\Lsc)}
\def\caseII{\TopfVW(\Asc,\Asc,\Asc,\Asc,\Lhh,\Lhh,\Lhh,\Lhh)}
\def\caseCrossed{\TopfVX(\Asc,\Asc,\Lhh,\Lsc,\Lhh,\Lsc,\Lsc,\Lhh,\Lsc)}

\def\topoIIxtra{\TopfVWdot(\Asc,\Asc,\Asc,\Asc,\Lsc,\Lsc,\Lsc,\Lsc)}
\def\topoIVxtra{\TopfVBdot(\Asc,\Asc,\Asc,\Asc,\Asc,\Asc,\Lsc)}
\def\topoXIIxtra{\TopfVBBdot(\Asc,\Asc,\Asc,\Asc,\Lsc)} 
\def\topoIIlap{\TopfVWlap(\Aqu,\Aqu,\Asc,\Asc,\Lsc,\Lsc,\Lsc,\Lsc)}
\def\topoIVlap{\TopfVBlap(\Asc,\Asc,\Aqu,\Asc,\Asc,\Aaqu,\Lsc)}
\def\topoXIIlap{\TopfVBBlap(\Asc,\Asc,\Asc,\Asc,\Lsc)} 

\title{Automatic reduction of four-loop bubbles 
\thanks{Talk presented at RADCOR/Loops and Legs 2002, Kloster Banz, Germany.
}
\hfill {\normalsize MIT-CTP 3326} 
}

\author{Y.~Schr\"oder \address{Center for Theoretical Physics, 
MIT, Cambridge, MA 02139, USA} 
}

\begin{document}

\begin{abstract}
We give technical details about the computational strategy employed 
in a recently completed investigation of the four-loop QCD free 
energy. 
In particular, the reduction step from generic vacuum bubbles to 
master integrals is described from a practical viewpoint, for
fully massive as well as QED-type integrals.
\end{abstract}

\maketitle

%
\section{Introduction}
\label{se:introduction}

Vacuum integrals, i.e. integrals without external momenta
(often also called tadpoles or bubbles), 
constitute an important class of multi-loop Feynman integrals.
While the perturbative expansion of quantities like the free energy 
can be directly expressed in terms of vacuum integrals, 
they also serve as essential building blocks for many other 
computations, being the coefficient functions in asymptotic expansions
of diagrams with external legs, and encoding the ultraviolet 
behavior of multi-scale integrals.

A typical perturbative calculation proceeds in four conceptually
independent steps. 
First, all relevant diagrams including their 
combinatoric factors are generated. 
For an algorithm that does this for vacuum integrals, see \cite{sd}.
Second, the Feynman rules of the theory under consideration are
inserted, and the color and Lorentz algebra is performed. 
Since in general individual loop integrals are divergent, a regularization
scheme has to be adopted, the most practical one at present 
being dimensional regularization (DR). 
Third, linear relations between the regularized integrals are exploited,
to systematically reduce all integrals occurring in the computation to a 
small set of so-called master integrals. 
In the framework of DR, the most important class of relations
can be derived from integration-by-parts (IBP) identities \cite{int_parts}.
Fourth, the master integrals have to be evaluated, 
either, in some fortunate cases, fully analytically, 
or as an expansion in terms of the regularization parameter,
in which case -- and only here -- the number of dimensions $d$
has to be specified. 
For results on the 4-loop level, see \cite{laporta4loop} ($d=4-2\e$)
and \cite{aleksi} ($d=3-2\e$). 

At higher loop orders, it is inevitable to automate the above 
setup to a large degree. 
There exist many approaches to implement automated perturbative 
calculations, and this is not the place to give a comprehensive review
(see e.g. \cite{algreview}). 
Instead, it is the third of the above steps that we wish to elaborate 
on in this contribution.

A computer algebra system that is particularly well suited to cope with
the demands of higher order perturbative calculations is FORM \cite{jamv}.
While by no means mandatory to use, we have adopted it to implement
our algorithms, and hence we will indicate in a few places which
specific FORM commands turned out to be extremely helpful.

%
\section{Notation and general considerations}
\label{se:notation}

\begin{figure}[t]
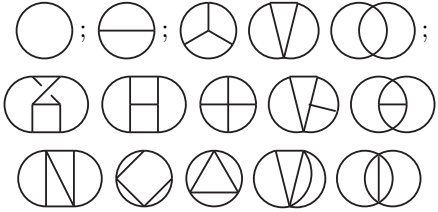

$$\one ; \two ; \threeM \threeV \threeB ; $$
$$\topoEX \topoI \topoII \topoIII \topoIV $$
$$\topoV \topoVI \topoVIII \topoIX \topoXII $$
\vspace{-10mm}
\caption[a]{\it The 1+1+3+10 generic vacuum topologies up to four loops.
The 0+1+2+6 factorized topologies are not shown here.}  
\label{fig:vactopos}
\vspace{-4mm}
\end{figure}

Consider the generic vacuum topologies of \fig\ref{fig:vactopos}.
In this intuitive graphical notation,
every line represents a propagator $(p_i^2+m_i^2)^{-a_i}$, with integer
power $a_i>0$, where the index $i$ labels the different lines with 
momenta $p_i$, which in turn can be expressed as a linear combination
of the $\l$\/ loop momenta $k_j$.
The vertices do not have any structure, except for assuring 
momentum conservation.
Each diagram can carry a nontrivial numerator structure, which in
the general case consists of powers of scalar products of the loop
momenta. 
At $\l$ loops, there are $\l(\l+1)/2$ different combinations 
$k_i\cdot k_j$. 

Let us distinguish three different representations of our integrals,
which naturally appear at various levels of the reduction process:
{\em generic integrals}, their {\em standard representations}, and 
the {\em master integrals}. 
The goal of step three is then to formulate the algorithms which
transform generic to standard to master integrals:
\ba \la{generic2master}
&&\hspace{-5mm}
\int_{k_{1\dots\l}}^{(d)} \fr{\prod_{1\le i\le j\le \l}\, 
(k_i\cdot k_j)^{b_{ij}}}
{\prod_{i}\,(p_i^2+m_i^2)^{a_i}} \nn
&&\hspace{-5mm}
\stackrel{\mbox{\footnotesize Filter}}\longrightarrow 
\int_{k_{1\dots\l}}^{(d)} 
\fr{\prod\,(k_i\cdot k_j)_{\mbox{\footnotesize irred.}}^{b_{ij}}}
{\prod_{i}\,(p_i^2+m_i^2)^{a_i}} 
\equiv \(\{a_i\},\{b_{ij},m_i\}\)
\nn
&&\hspace{-5mm}
\stackrel{\mbox{\footnotesize Tables}}=\,
\sum_j c_j(d)\, \mbox{Master}_j^{(d)}(\{m_i\})
\ea

Above, the label 'Filter' symbolizes a collection of low-level routines,
whose main action is to complete squares in the numerator and cancel against 
propagators such that only irreducible numerators remain. 
At this point, it is possible to represent an $\l$\/-loop vacuum
integral by a list of $\l(\l+1)$ non-negative numbers 
$(\{a_i\},\{b_{ij},m_i\})$, the first half of them collecting the
powers of propagators $a_i$, while the second half
contains either the power of an irreducible numerator
(if the corresponding $a_i$ is zero) or the mass of the line.
Furthermore, at this step 
equivalent topologies are re-labeled in a unique way 
by shifting the loop momenta, i.e. assigning a characteristic
pattern of zeroes among the $a_i$ to each topology of \fig\ref{fig:vactopos}.

In the remainder, we will specialize on two different general 
classes of vacuum diagrams. First, we will consider all lines
to have the same mass, $m_i=m$. This class of integrals is 
useful when computing infrared-safe quantities like renormalization
coefficients, in which case the infrared sector of individual diagrams
can be regulated by introducing masses into massless propagators.
Second, we will allow for all $m_i$ to be either zero or $m$, 
with the restriction that the number of massive lines at each
vertex be even. This includes theories like QED and gauge+Higgs
models, whence we call this class 'QED-like'. 

The label 'Tables'  in \eq\nr{generic2master} symbolizes a lookup 
in a database, which contains
the necessary relations in a tabulated form. 
These tables are the main ingredient of the reduction step, 
and their organization and generation, 
which systematically exploits IBP identities, 
will be described in more detail below.

The intermediate step of applying the 'Filter' algorithms not only
serves the purpose of allowing for a fairly compact representation
of the integral, but can also be used 
to keep the number of entries in the database, the memory requirements,
and the CPU time needed for their derivation, in manageable bounds.
To this end, we found it advantageous to add further routines
to the 'Filter' package: 
\begin{itemize}
\addtolength{\itemsep}{-2mm}
\item Early detection of zeroes: massless (sub-) tadpoles are zero in DR,
as are integrals whose integrand does not depend on one of the loop 
momenta. 
\item Symmetrization of the integrand: use the full symmetry group of
the corresponding topology to order the list, and hence enable early
cancellations in big expressions.
\item Decouple scalar products involving the loop momentum of a 
factorized one-loop tadpole: 
$\int_k\! \fr{k_{\mu_1}\dots k_{\mu_n}}{(k^2+m^2)^a}$ vanishes for odd $n$
and is proportional to a totally symmetric combination of metric tensors
$g_{\{\mu_1\mu_2}\dots g_{\mu_{n-1}\mu_n\}}$. The FORM function {\tt dd\_}
is perfectly suited for this symmetrization.
This eliminates the need to derive relations for 8 (out of 9, the 9th 
being the two-loop $\times$ two-loop case) of the
factorized topologies, since after decoupling the numerator, factorization
into scalar vacuum integrals of the type of \eq\nr{generic2master} 
is complete.
\item Reduce powers of factorized one-loop tadpoles to one:
{\footnotesize
$$\int\! \fr{{\rm d}^dk}{(k^2+m^2)^{a+1}}=-\fr{d-2a}{2am^2}
\int\! \fr{{\rm d}^dk}{(k^2+m^2)^a}$$}
\item Employ the 'triangle relation' \cite{int_parts}: 
for integrals involving massless lines, 
this helps to reduce the number of different topologies that
have to be treated in the database considerably.
\end{itemize}
Another potentially useful routine, which we have however not implemented, 
would be to use $T$\/-operators \cite{tarasovT} in order to trade 
{\em all}\/
numerator structure for higher dimensions of the integral measure,
hence also immediately decoupling the factorized (two-loop $\times$ 
two-loop)-topology.

One more practical note:
To not miss cancellations, it is
important to have a unique representation for coefficients. 
Partial fractioning of terms like $\fr{d}{d-a}$ and 
$\fr1{d-a_1}\,\fr1{d-a_2}$ helps here, ensuring the coefficients $c_i(d)$ 
to be a sum of powers of simple poles $\fr{1}{d-a}$ and powers of $d$.

In principle however, all these further relations are redundant since
they would be automatically covered by the IBP identities.
As mentioned above, their sole purpose is to optimize the 
derivation of relations among the integrals, to be discussed next. 

%
\section{Reduction}
\label{se:reduction}

Integration by parts relies on the fact that an integral over a 
total derivative of
any of the loop momenta vanishes in dimensional regularization.
For the case of vacuum integrals, which we are interested in here, 
the IBP identities read
\be \la{eq:generalPI}
0=\int\fr{{\rm d}^dk_{1\dots\ell}}{(2\pi)^{\ell d}} \,
\6_{p_\mu} q_\mu \, \(\{a_i\},\{b_{ij},m_i\}\) \;,
\ee
where $p,q\in\lb k_1,\dots,k_\l\rb$ cover all $\l^2$ 
different $\l$\/-loop identities,
and we have made use of the standard representation introduced above. 

There are two possible general strategies
implementing the IBP identities to find relations useful for
reducing the integrals from their standard representation to 
master integrals.

The first strategy is to derive general relations, valid for 
symbolic list-entries. These general symbolic relations 
can then be applied repeatedly to any integral of the 
specified class, no matter how large the powers are, 
to achieve the reduction.
In practice however, it turns out that it is quite an art
to shuffle IBP identities for integrals with symbolic indices
such as to obtain useful reduction relations.
In absence of a generic algorithmic
formulation, it involves extensive handwork, and typically there
are many special cases to be considered when pre-factors vanish 
at special parameter values. 
At lower loop orders, there are complete solutions,
see e.g. \cite{tarasov2loop} for 
two-loop two-point functions with general masses,
or \cite{vac3loop} for three-loop vacuum integrals with one mass. 

The second strategy, 
nowadays constituting the mainstream of higher-loop
computations, is a more brute-force approach, which however has the
huge advantage of being perfectly suited to be completely automated.
The main idea is to write down IBP identities for specific
values of the indices. Introducing a lexicographic ordering
among the integrals \cite{laportaAlgorithm},
it is then possible to solve every single one of the IBP identities
for the 'most difficult' integral occurring.
By starting from simple topologies (low number of lines), one
systematically generates relations which express 'difficult' integrals
in terms of 'simpler' (in the sense of the ordering) ones. 
Solving an adequate set of fixed-index
IBP relations, it is possible to express every integral of interest
in terms of a few simple ones, ultimately the master integrals. 

The reason why the second strategy is sufficient for most 
computations is that in practice, one does not meet the most
general integrals, but only a subset, typically characterized by
an upper cutoff on the sum of indices. Indeed, dealing with a 
concrete model like QCD, knowledge of the vertex and propagator
structure allows to constrain the set of possible indices 
$\(\{a_i\},\{b_{ij},m_i\}\)$, hence rendering the search-space 
to be covered with IBP identities finite.

\begin{figure}[t]
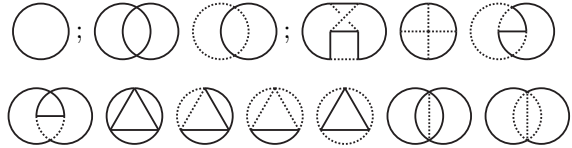

\vspace{-3mm}
$$\one ; \threeB \threeQED ; \caseCrossed \caseII \caseIII$$
$$\caseIV \caseVII \caseVIII \caseIX \caseX \caseXIII \caseXIV$$
\vspace{-10mm}
\caption[a]{\it The 1+0+2+10 master integrals of QED type, up to four 
loops. Full lines carry a mass $m$, dotted lines are massless. 
All numerators are 1, all powers of propagators are 1.
Note that there is no two-loop representative needed.}  
\label{fig:qedints}
\vspace{-4mm}
\end{figure}

Building up the relations proceeds as follows:
\begin{itemize}
\addtolength{\itemsep}{-2mm}
\item Pick a list of indices $\(\{a_i\},\{b_{ij},m_i\}\)$ that is 'simple',
typically meaning a low number of loops, a low number of different
lines, a low number of extra powers on the propagators, a low
number of powers on the irreducible numerators. 
In FORM, these lists are most naturally represented by sparse tables.
\item Generate the first of the IBP identities. 
\item Call 'Filter' to transform
the resulting sum of integrals to the standard representation.
\item Label the 'most difficult' integral, according to the lexicographic
ordering. 
A global operation like that became possible
with the introduction of '$\$$\/-variables' in FORM v3. 
\item Invert its coefficient and multiply it into the equation. 
We do so only if we can factorize the coefficient into terms which 
are linear in $d$, to preserve the generic structure of coefficients. 
While (at present) there is no factorization algorithm in FORM,
we implemented one by 'guessing' zeroes, utilizing the fact that
since the coefficients are generated by IBP, most of them have
factors $(nd\pm a)$ where $n$ is not bigger than the number of loops,
and $a$ is an integer of moderate size.
To check that no relations are missed when factorization
fails, it is useful to keep track of those cases and check in 
the end.
\item Bring the most difficult integral (having coefficient $1$)
to the left-hand-side, taking the generated equation as a definition.
In FORM, this is done by the {\tt fill} statement.
\item Take the next IBP identity, repeat the above steps. 
Increase list-indices. Repeat...
\item Write the relations found to disk in intervals. 
Large intervals ensure a high degree of re-substitution
(of relations for integrals that are found later but that appeared 
on the right-hand-sides earlier), 
but are risky when the program execution crashes. 
\end{itemize}
Solving the IBP relations one by one like described above seems
to be simpler than solving large systems of linear equations at once.
In the end, it might be advantageous to re-substitute relations,
which is possible by re-loading sets of relations into
memory and re-writing them to disk. 

In the end, one has to check whether the set of generated identities
is sufficiently large to achieve a reduction of all integrals
occurring in the physics problem at hand. 
While a first educated guess on the maximum powers needed
can be obtained by scanning the terms to be calculated after application
of the 'Filter' package, it might be necessary to enlarge the
set of relations in further runs. To this end, the {\tt tablebase}
statement of FORM, implemented in version {3.1}, allows for a
good control over large amounts of data in the form of tables
and table elements.

\begin{figure}[t]
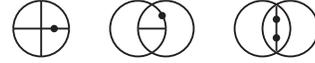

$$\topoIIxtra \quad \topoIVxtra \quad \topoXIIxtra $$
\vspace{-10mm}
\caption[a]{\it The 0+0+0+3 fully massive master integrals, 
in addition to those 1+1+3+10
of \fig\ref{fig:vactopos}, taken at powers and numerators 1.
A dot on a line means it carries an extra power.}  
\la{fig:massints}
\vspace{-4mm}
\end{figure}

The resulting master integrals are depicted in \fig\ref{fig:qedints}
for the 'QED-like' case, and in 
\figs\ref{fig:massints},\ref{fig:vactopos} 
for the fully massive case. 

%
\section{Master integrals}
\label{se:masterintegrals}

\begin{figure*}[t]
\vspace{-2mm}
{\footnotesize
\begin{eqnarray*}
40(2d\!-\!3)(3d\!-\!4) \topoXIIlap (k_1\!\cdot k_2)^2 &=& 
900 \topoXIIxtra 
-(1300-733d+57d^2) \topoXII 
-5(16-43d+21d^2) \topoXV 
\\[1ex]
40(3d\!-\!8) \topoIVlap (k_1\!+\!k_2)^2 &=& 
-240 \topoIVxtra 
-40(5d-16) \topoIV 
-\fr{250}{d-3} \topoXIIxtra \\&&{}
-\fr{2(2d-5)(6d-13)}{d-3} \topoXII 
+30(3d-8) \topoIX \\&&{}
+80(d-2) \one \!\!\(\!\! \threeV + \threeB \!\!\) 
-\fr{25(d-2)^2}{d-3} \topoXIV 
\\[1ex]
2(d\!-\!3) \topoIIlap (k_1\!\cdot k_2) &=& 
6 \topoIIxtra 
+(d-4) \topoII
+12 \topoIVxtra 
+2(2d-7) \topoIV 
+\fr5{d-3} \topoXIIxtra \\&&{}
-\fr{(2d-7)(2d-5)}{5(d-3)} \topoXII  
+(3d-8) \topoIX 
-2(d-3) \topoVIII \\&&{}
-(d-2) \topoXI 
+\fr{(d-2)^2}{2(d-3)} \topoXIV 
\end{eqnarray*}}
\vspace{-10mm}
\caption[a]{\it 
Relations for a basis conversion from the set of massive masters found in 
\cite{laporta4loop} to our notation.}
\label{fig:basisconv}
\vspace{-4mm}
\end{figure*}

Once the reduction algorithm 'stops', 
are we guaranteed to arrive at the desired {\em minimal} set of
master integrals? 
If we had followed the path of deriving generic reduction relations,
valid for symbolic indices, the answer would be yes. 
For the implementation in terms of specific indices, one can however 
not be absolutely sure not to miss a relation which would only be detected
when increasing the upper cutoff on indices of the integrand. 
For most practical purposes it might already be sufficient
to work with an incomplete, but small, basis.

In the case of gauge theories, it is also amusing to watch the 
gauge-parameter dependence as an indicator of how 'close' one
is to the minimal set, since in a full reduction gauge-parameter 
dependent terms cancel at an algebraic level,
in $d$ dimensions, before evaluating the master integrals.

The basis of master integrals is of course not unique, but depends on the 
actual choice of the lexicographic ordering.
While we label an integral with unit numerator as 'simpler' than
one with increased powers on the lines, one could as well have
put priority on always reducing powers of propagators to one.
The latter choice was adopted in a recent paper \cite{laporta4loop},
where the three master integrals of \fig\ref{fig:massints} are replaced
by ones of equivalent topology, but with irreducible numerators.
Consequently, there must be linear relations between the two choices of
basis, valid analytically in $d$ dimensions.
From our tables, we simply read them off, see \fig\ref{fig:basisconv}.

%
\section{Discussion}
\label{se:discussion}



We did not comment on the problem of so-called spurious poles here. 
Spurious poles are singular pre-factors, which can occur in the
reduction relations.
They are difficult to avoid in
general, if one is not willing to specify the dimension yet 
in the reduction process. However, in our four-loop computation
of the QCD free energy, we treated them after the reduction was
performed successfully, by changing basis with the help of the
tables.

In principle, the package at hand can be used for other calculations
requiring a four-loop reduction of massive vacuum bubbles. 
One such application would be the re-evaluation of the QCD beta function.

\vspace{3mm}
\noindent
{\bf Acknowledgments:} It is a pleasure to thank K.~Kajantie,
M.~Laine and K.~Rummukainen for an enjoyable collaboration on the
matter presented in the above. This work was supported in parts 
by the DOE, under Cooperative Agreement no.~DF-FC02-94ER40818.

%

\end{document}